\documentclass{PoS}

\title{Semileptonic B decays}

\ShortTitle{Semileptonic B decays}

\author{Giulia Ricciardi
\\Dipartimento di Fisica, Universit\'a degli Studi di Napoli Federico II, Via Cintia, 80126 Napoli, Italy and
INFN, Sezione di Napoli, Via Cintia, 80126 Napoli, Italy\\
        E-mail: \email{giulia.ricciardi@na.infn.it}}

\abstract{We review recent progress in exclusive  $B \to D^{(\ast)}$ and inclusive $B \to X_u$    semileptonic B decays.}

\FullConference{Xth Quark Confinement and the Hadron Spectrum,\\
		October 8-12, 2012\\
		TUM Campus Garching, Munich, Germany}

\def\beq{\begin{equation}}
\def\be{\begin{equation}}
\def\eeq{\end{equation}}
\def\ee{\end{equation}}
\def\bea{\begin{eqnarray}}
\def\eea{\end{eqnarray}}

\newcommand{\gsim}{\mbox{${~\rightarrowise.25em\hbox{$>$}\kern-.70em
\lower.25em\hbox{$\sim$}~}$}}
\newcommand{\lsim}{\mbox{${~\rightarrowise.25em\hbox{$<$}\kern-.70em
\lower.25em\hbox{$\sim$}~}$}}

\def\ra{\rightarrow}
\def\D0{D\O }

\begin{document}

\section{Introduction}

In the last years, detailed studies of semi-leptonic decays of $B$ mesons have been made  possible by 
 the vast samples of $B$ mesons recorded at CLEO and, especially, at the B factories. Also  \D0 and LHCb have helped,  collecting data on
$B$ mesons decays into excited $D$ meson states and $B_s$ decays.
Ultimate precision should be reached  by future SuperB and Belle II  facilities.

 Semi-leptonic decays   play a critical role in
 the analysis of the unitarity triangle.
 Information  about the CKM matrix elements $ |V_{cb}| $ and
$|V_{ub}|$ is provided by  exclusive and inclusive processes driven by $ b \ra c(u)\, l \nu_l$  \cite{Ricciardi:2012pf, Ricciardi:2012dj}.
We can compare, as an example, with  leptonic decays $ B^+  \rightarrow l^+ \nu $.
The decay rates depend on the product of the  modulus of the  $|V_{ub}|$  and of the decay constants $f_B$, therefore the  determination of
the CKM element requires
the theoretical knowledge  of  $f_B$. 
In addition, the comparison between experimental and  theoretical results  presents some hardships.
Experimental disadvantages come from the helicity suppression of the leptonic decays and  the comparatively high rate of the  process $  B^+ \rightarrow l^+ \nu \, \gamma$.
For decays into tauons,  these effects
appear to be small, while  the helicity suppression is mostly broken.
Moreover, the existence of a charged Higgs boson (or any other charged
object beyond the Standard Model) would easily modify the decay rates.
Exclusive semileptonic decays can be   called to the rescue by supplying  complementing informations.
They are not helicity suppressed, still they retain the possibility of theoretical simplification by factorizing the leptonic current. They are tree level dominated and new physics is generally not expected (at least to a high
degree of accuracy).

 Here we review recent progress in exclusive $B \to D^{(\ast)}$ and inclusive $B \to X_u$  semileptonic B decays.

\section{Exclusive $B \to D^{(\ast)}$   decays}

The  differential ratios for the semileptonic  decays 
$ \bar{B}\rightarrow D \, l \, \bar{\nu}$ and $ \bar{B}\rightarrow D^\ast \, l \, \bar{\nu}$ 
depend on  several unknown form factors, that are constrained and reduced 
in the heavy quark limit.
Neglecting  the charged lepton and neutrino masses, we can write
\bea
\frac{d\Gamma}{d \omega} (\bar{B}\rightarrow D\,l \bar{\nu})  &=&  \frac{G_F^2}{48 \pi^3}\,  (m_B+m_D)^2 \, m_D^3
(\omega^2-1)^{3/2}\,  |V_{cb}|^2 ({\cal G}(\omega))^2 \nonumber \\
\qquad\frac{d\Gamma}{d \omega}(\bar{B}\rightarrow D^\ast\,l \bar{\nu})
&=&  \frac{G_F^2}{48 \pi^3}   (m_B-m_{D^\ast})^2 m_{D^\ast}^3   (\omega^2-1)^{1/2}  \chi (\omega)  |V_{cb}|^2 
({\cal F}(\omega))^2\label{excl1} \eea
Here ${\cal F}(\omega)$ and ${\cal G}(\omega)$ are the form factors in the heavy quark limit,
 $\chi (\omega)$ is a known phase space
factor and  $\omega= v_B \cdot v_{D^{(\ast)}}= (m_B^2+m_{D^{(\ast)}}^2 -q^2)/2 m_B m_{D^{(\ast)}} $ is  the product of the heavy quark velocities  $v_B= p_B/m_B$ and $v_{D^{(\ast)}}= p_{D^{(\ast)}}/m_{D^{(\ast)}}$. 
Let us observe that outside the   heavy quark limit ${\cal F}(\omega)$
contains a combination of three form factors (one vector
and two axial vectors) related to the three $D^\ast$ helicity
states.
 In the $ B$ rest frame, $\omega$ 
 corresponds to the energy of $D^{(\ast)}$ normalized by its mass, that is to $\omega =E_{D^{(\ast)}}/m_{D^{(\ast)}}$. 
 The values of $\omega$  are constrained by kinematics: $ 
\omega \geq 1$, with largest value $ \omega \simeq 1.5$. 
At the nonrecoil point,  $\omega
= 1$,  heavy quark symmetries play a useful role
in setting the normalization ${\cal F}(1)={\cal G}(1)$; this point, however, is not directly accessible experimentally due to the kinematic suppression
in Eqs. (\ref{excl1}).
The main theoretical problem is  the non-perturbative evaluation of the form factors. By exploiting Eqs.  (\ref{excl1}), the product $|V_{cb}| \times $ form factor can be extracted by experimental fits to the differential decays rates. 

At present the $ \bar{B}\rightarrow D^\ast \, l \, \bar{\nu}$   decay  is measured with a better
accuracy than the $ \bar{B}\rightarrow D\, l \, \bar{\nu}$  decay, therefore the exclusive determination of $| V_{cb}|$ relies
on the theoretical non-perturbative  results for the form factor  ${\cal F}$. 
 The most recent  the Heavy Flavor Averaging
Group  (HFAG) experimental  fit \cite{Amhis:2012bh} gives
\beq  |V_{cb}| |{\cal F}(1)| = (35.90 \pm 0.45 ) \times   10^{-3} \label{VcbexpF1} \eeq 
The only unquenched result available
 \cite{Bernard:2008dn, Bailey:2010gb} gives 
\beq {\cal F}(1)
=0.908\pm 0.017 \label{F11}  \label{VcbexpF2} \eeq 
including  the enhancement factor 1.007, due to the
electroweak corrections to the four-fermion operator mediating
the semileptonic decay.
By taking Eqs.  (\ref{VcbexpF1}) and   (\ref{F11}),  the latest  value of $|V_{cb}|$   reads \cite{Amhis:2012bh}
  \beq
|V_{cb}| = (39.54 \pm 0.50_{\mathrm{exp}} \pm 0.74_{\mathrm{th}}) \times 10^{-3} \label{lattice11}
\eeq 
 where the errors come from experiment and QCD lattice calculations, respectively.
The
${\cal F}(1)$  form factor has also  been  calculated  recently using zero recoil sum rules. The quoted value \cite{Gambino:2010bp, Gambino:2012rd}, including full $\alpha_s$ and up to $1/m_5^2$, 
is
\beq {\cal F}(1) = 0.86 \pm 0.02 \label{gmu} \eeq
By using the same  Eq.~(\ref{VcbexpF1}) in combination with Eq.~(\ref{gmu}), one obtains
\beq
|V_{cb}| =  (41.6\pm 0.6_{\mathrm{exp}}\pm 1.9_{\mathrm{th}}) \times 10^{-3} 
\label{VCBF1}
\eeq
an higher value, in average and in theoretical error, than the one given by the lattice calculation.

Let us compare the previous determinations with the $|V_{cb}| $ value extracted from  $ \bar{B}\rightarrow D \, l \, \bar{\nu}$ decays. The most recent  unquenched calculation dates back to 2005  \cite{Okamoto:2004xg} and gives, after 
 correcting by the usual factor of 1.007 \beq {\cal G}(1) = 1.074 \pm 0.024 \eeq
which, using
the latest HFAG average \cite{Amhis:2012bh}, that
includes older Aleph, CLEO and Belle measurements, as well as the new 2008-2009 BaBar  data, becomes
\beq |V_{cb}| | {\cal G}(1)| = (42.64 \pm 1.53 ) \times
10^{-3} \eeq  The resulting  estimate is
 \beq |V_{cb}| = (39.70 \pm 1.42_{\mathrm{exp}} \pm 0.89_{\mathrm{th}})
\times  10^{-3} \label{lattice2} \eeq 
 in good agreement with the lattice determination from  $ \bar{B}\rightarrow D^\ast \, l \, \bar{\nu}$  Eq. (\ref{lattice11}).
However, the experimental error  is more than twice larger.
An alternative lattice determination,  currently available only in the quenched approximation,  consists in calculating  the form factor normalization directly  at values $\omega >1$,
 avoiding the large extrapolation to $\omega=1$ and thus reducing the model dependence  \cite{de Divitiis:2007ui, deDivitiis:2007uk}. 
This approach, 
by using 2009  BaBar  data \cite{Aubert:2009ac}, gives a slightly higher value than the lattice result (\ref{lattice2}),  that is
 \beq |V_{cb}| = (41.6 \pm 1.8 \pm 1.4
\pm 0.7_{FF} ) \times  10^{-3} \label{lattunq} \eeq
The errors are  statistical, systematic and due to the theoretical uncertainty in the form factor $ {\cal G}$, respectively.
The most recent non lattice calculation combines the heavy quark expansion with a  ''BPS" expansion \cite{Uraltsev:2003ye}
giving  in this limit
 \beq {\cal G}(1) =1.04 \pm  0.02 \eeq
With such estimate one finds \cite{Beringer:1900zz}
\beq
|V_{cb}| =  (40.7 \pm 1.5_{\mathrm{exp}} \pm 0.8_{\mathrm{th}}) \times 10^{-3} 
\eeq
in agreement, within the errors, with both lattice 
determinations  (\ref{lattice2})  and (\ref{lattunq}).

Until a few years ago,  only decays when the final lepton was an electron or a muon had been observed.
When the lepton mass is different from zero, the relations (\ref{excl1})
do not hold anymore, and 
experimental measurements of
 $B \to D^{(\ast)} \tau \nu$  decays are no more sensitive to one form factor  only, even in the heavy quark limit. It has been recently observed that 
 $B$  leptonic and semileptonic decays with a $\tau$ in the final state exhibit a tension of order $2 \sim 3 \sigma$  between experimental and theoretical results.
Branching fraction measurements of $B^- \to \tau^- \bar \nu_\tau$  decay differ from the theoretical predictions based on the standard model fits, and the value of $|V_{ub}|$ represents the main source of theoretical uncertainty.
BaBar collaboration has recently published 
results of their measurements of $B \to D^{(\ast)} \tau \nu$  branching fractions normalized to the
corresponding $B \to D^{(\ast)} l \nu$ modes (with $l=e , \mu$) by using  the full BaBar data sample \cite{Lees:2012xj}
\beq
R(D) = \frac{ {\cal{B}}(B \to D \tau \nu)}{{\cal{B}}(B \to D l \nu)}= 0.440 \pm 0.072 \qquad
R(D^\star) =\frac{ {\cal{B}}(B \to D^{\ast} \tau \nu)}{{\cal{B}}(B \to D^\ast l \nu)}= 0.332 \pm 0.03 
 \label{BaBartau}
\eeq
where the statistical and systematic uncertainties have been combined in quadrature.
The results  (\ref{BaBartau})  have  been  compared with the SM predictions, finding  $  R (D)_{\mathrm SM} =   0.297 \pm
0.017$ and   $ R (D^\ast)_{\mathrm SM} = 0.252\pm 0.003 $, averaged
over electrons and muons.
They  respectively exceed the SM predictions
by $2.0$ and $2.7 \sigma$, corresponding to a combined discrepancy at the $3.4 \sigma$ level.
A recent  phenomenological approach questions the SM determination for $R(D)$ giving 
 $R(D)=0.31 \pm 0.02$ \cite{Becirevic:2012jf} and  a similar result,  $R(D)=0.316 \pm 0.012 \pm 0.007$, is found 
in  a recent  full, 2+1 flavor lattice QCD calculation 
\cite{Bailey:2012jg}: 
both analysis reduce the significance of the discrepancy
below $2\sigma$.
The values for $ R(D)$ given by
 Belle \cite{Adachi:2009qg} agree with those of BaBar, but have larger
uncertainties. Current experimental measurements of
$R(D)$ are statistics-limited, so the luminosities available
at future flavor facilities should enable significant improvements.
If the experimental results for  $ R(D^{(\ast)})$ will be confirmed, 
they might point to new physics effects in  semitauonic
B decays. 
It has been excluded that the  excess in Eqs. (\ref{BaBartau}) over the SM  can be explained  within  type II two Higgs doublet model \cite{Lees:2012xj} and in the context of the MSSM with MFV  \cite{Altmannshofer:2012ks}. Instead, it is deemed  possible in other SM extensions, e.g. in  type III \cite{Crivellin:2012ye} and Aligned \cite{Celis:2012dk} two Higgs doublet models,
 by 
adopting effective Lagrangians
\cite{Fajfer:2012vx, Fajfer:2012jt, Datta:2012qk},  and so on.

\section{Inclusive $b \to u$   decays}

In inclusive $ B \rightarrow X_q  l  \nu$ decays, we sum over all possible final states $X_q$, no matter if single-particle  or  multi-particle states.
Since  inclusive decays  do not depend on the details of final state, quark-hadron duality is generally assumed.
In most of the phase space,  long and short distance dynamics are factorized by means of  the heavy 
quark expansion.
However, the  phase space region includes a  region of singularity, also called endpoint or threshold region, plagued by the presence 
 of large double (Sudakov-like)  perturbative  logarithms at all orders in the strong coupling\footnote{for
 theoretical aspects of threshold
resummation in $B$ decays
see e.g.  \cite{Aglietti:2002ew, Aglietti:2005mb, Aglietti:2005bm, Aglietti:2005eq, Aglietti:2007bp, DiGiustino:2011jn}.}.
For  $b \rightarrow c$ semileptonic decays, the effect of the small region of singularity is not very important; in addition,  corrections are not expected  as singular as in the $ b \rightarrow u$ case, being  cut-off by the charm mass.

The analysis of 
inclusive charmless semileptonic decays   is currently employed to determine the CKM parameter $|V_{ub}|$, which
 plays a crucial role in the study  of
the unitarity constraints and related questions.
 In principle,  the method of extraction of $|V_{ub}|$  from inclusive $ \bar  B \rightarrow X_u  l \bar \nu_l$  decays  follows in the footsteps of the $|V_{cb}|$ determination from $ \bar  B \rightarrow X_c  l \bar \nu_l$ , that employs the  OPE expansion 
in  the systematic
framework provided by the heavy quark effective theory.
However, in $ \bar  B \rightarrow X_u  l \bar \nu_l$ decays,  the copious background from the
$ \bar B \rightarrow X_c l \bar \nu_l$  process, which has a rate about 50 times higher, stands in the way, and 
the experimental sensitivity is higher in restricted regions of phase space,
where such background  is  highly suppressed by kinematics. They include the threshold  region, where the previous approch fails, and new
theoretical issues need to be addressed.
A lot of effort has  been devoted to enlarge
the  experimental range, so to reduce
on the whole  the weight of the endpoint region.
 Latest results by Belle \cite{Urquijo:2009tp}
 access about the $ 90$\% of the $ \bar B \rightarrow X_u  l \bar \nu_l$ phase space, claiming an overall uncertainty of 7\% on $|V_{ub}|$.
A similar portion of the phase space is covered also by the more recent BaBar analysis \cite{Lees:2011fv}.
On the theoretical side several approaches have been devised 
to analyze data in the threshold region,  with differences
in treatment of perturbative corrections and the
parameterization of nonperturbative effects.

Experimental results are provided by CLEO, BaBar and Belle;  the B factories have provided the largest and more recent amount of data. 
The average values for $|V_{ub}|$ can be extracted  from the partial branching fractions, adopting
a specific theoretical framework and taking into account
 correlations among the various measurements
and theoretical uncertainties.
The averaging procedure
has been developed by HFAG \cite{Amhis:2012bh}, that has also checked  the consistency among  inputs of the different theoretical frameworks.
In
Table \ref{phidectab2} we list the  HFAG averages,  that 
are determined by comparing the measured partial branching fractions in selected regions of phase-space to theoretical calculations of the corresponding rates.
We have listed all the QCD theoretical calculations that take into account  the whole set of experimental results, or most of it, starting from 2002 CLEO data \cite{Bornheim:2002du}.
Very roughly speaking, the 
 BLNP approach
by Bosch, Lange, Neubert, and Paz \cite{Lange:2005yw} and the
GGOU one by Gambino, Giordano, Ossola
and Uraltsev \cite{Gambino:2007rp}  differ in their treatment of Sudakov double logarithms resumming and 
subleading shape function modeling.
The DGE, the
dressed gluon exponentiation, by Andersen and Gardi \cite{Andersen:2005mj, Gardi:2008bb} and the ADFR approach, by Aglietti, Di Lodovico, Ferrara, and Ricciardi,
\cite{Aglietti:2004fz, Aglietti:2006yb, Aglietti:2007ik}  start analyzing the singularities of the
perturbative expansion
in moment space; in the ADFR approach, an effective running coupling is introduced to help estimating nonperturbative corrections. 
The results  listed in  Table \ref{phidectab2}  are consistent within the errors, but  the theoretical uncertainty among determinations can reach 10\%.
Other solutions have been proposed in \cite{Bauer:2001rc, Leibovich:1999xf, Ligeti:2008ac}.

Notwithstanding all the experimental and theoretical efforts, 
the values of $|V_{ub}|$ extracted from inclusive decays maintain about two $\sigma$ above the values given by exclusive determinations. The most recent BaBar estimate from exclusive decays yields  
$
|V_{ub}| = (3.25 \pm 0.31) \times 10^{-3} \label{exclus} $  \cite{Lees:2012vv} and it is determined
from the simultaneous fit to the experimental data
and the lattice  theoretical predictions. It is compatible
with the Belle result \cite{Ha:2010rf} of  $
|V_{ub}| = (3.43 \pm 0.33) \times 10^{-3}  $.
Also indirect fits prefer a lower value of $|V_{ub}|$, such as 
  $ |V_{ub}|=  (  3.62 \pm 0.14 ) \times 10^{-3}  $ 
by the UTfit Collaboration \cite{Tarantino:2012mq}.

\begin{table}[t]
\centering
\vskip 0.1 in
\begin{tabular}{l c} \hline \hline
{\it \bf Theory}  &   $|V_{ub}| \times 10^{3}$ \\
\hline
\hline
BLNP  & $ 4.40 \pm 0.15^{+0.19}_{-0.21}  $\\
\hline
DGE   & $4.45 \pm 0.15^{+ 0.15}_{- 0.16}$\\
\hline
ADFR   & $4.03 \pm 0.13^{+ 0.18}_{- 0.12}$\\
\hline
GGOU   & $4.39 \pm  0.15^{ + 0.12}_ { -0.20} $\\
\hline
\hline
\end{tabular}
\caption{ Comparison of  inclusive  determinations of $|V_{ub}|$ \cite{Amhis:2012bh}. }
\label{phidectab2}
\end{table}

\end{document}